\documentclass[runningheads,a4paper]{llncs}
\usepackage{amssymb}
\usepackage{graphicx}
\usepackage{url}

\usepackage{amsfonts}
\usepackage{url}
\usepackage[boxed,vlined,ruled,linesnumbered]{algorithm2e}
\usepackage{lhelp}

\begin{document}

\mainmatter

\title{Using the Regular Chains Library to build cylindrical algebraic decompositions 
by projecting and lifting.}  
\titlerunning{Using the \textsc{RegularChains} Library to build CADs by projection and lifting.} 
\author{Matthew England \and David Wilson \and Russell Bradford \and James H. Davenport}
\authorrunning{England-Wilson-Bradford-Davenport}
\institute{
University of Bath, UK\\
\email{ \{M.England, D.J.Wilson, R.J.Bradford, J.H.Davenport\}@bath.ac.uk},\\ 
}
\maketitle

\vspace*{-10pt}

\begin{abstract}
Cylindrical algebraic decomposition (CAD) is an important tool, both for quantifier elimination over the reals and a range of other applications.  Traditionally, a CAD is built through a process of projection and lifting to move the problem within Euclidean spaces of changing dimension.  Recently, an alternative approach which first decomposes complex space using triangular decomposition before refining to real space has been introduced and implemented within the \textsc{RegularChains} Library of \textsc{Maple}.  
We here describe a freely available package \texttt{ProjectionCAD} which utilises the routines within the \textsc{RegularChains} Library to build CADs by projection and lifting.  
We detail how the projection and lifting algorithms were modified to allow this, discuss the motivation and survey the functionality of the package.
\end{abstract}

\section{Introduction} 
\label{SECTION:Intro}

A \textit{cylindrical algebraic decomposition} (CAD) is: a \textit{decomposition} of $\mathbb{R}^n$, meaning a collection of cells which do not intersect and whose union is $\mathbb{R}^n$; \textit{cylindrical}, meaning the projections of any pair of cells with respect to a given variable ordering are either equal or disjoint; and, \textit{(semi)-algebraic}, meaning each cell can be described using a finite sequence of polynomial relations.
CAD is best known for quantifier elimination over the reals, but has also found diverse applications such as motion planning \cite{WDEB14} and reasoning with multi-valued functions \cite{DBEW12}.

The \textsc{RegularChains} Library \cite{RC} in \textsc{Maple} contains procedures to build CAD by first building a \emph{complex cylindrical decomposition} (CCD) of $\mathbb{C}^n$ using triangular decomposition by regular chains, then refining to a CAD of $\mathbb{R}^n$.
The core algorithm was developed in \cite{CMXY09} with improvements detailed in \cite{CM12b} and \cite{BCDEMW14}.

These CAD algorithms are in contrast to the traditional approach of projection and lifting followed since Collins' original work \cite{Collins1975}.  Here, a \textit{projection} phase repeatedly applies an operator to a set of polynomials (starting with those forming the input) each time producing another set in one fewer variables.  Then the \textit{lifting} phase builds CADs of $\mathbb{R}^i, i=1 \dots n$.
$\mathbb{R}$ is decomposed into points and intervals corresponding to the real roots of the univariate polynomials.  $\mathbb{R}^2$ is decomposed by repeating the process over each cell in $\mathbb{R}^1$ using the bivariate polynomials at a sample point.  The output over each cell consists of {\em sections} (where a polynomial vanishes) and {\em sectors} (the regions between) which together form a  {\em stack}.  The union of these stacks gives the CAD of $\mathbb{R}^2$ and the process is repeated until a CAD of $\mathbb{R}^n$ is produced.  
Collins defined the projection operator so the CAD of $\mathbb{R}^n$ produced using sample points this way could be concluded \textit{sign-invariant} for the input polynomials: each polynomial has constant sign on each cell.  
The key tool in the proof was showing polynomials to be \emph{delineable} in a cell, meaning the zero set of individual polynomials are disjoint sections and the zero sets of different polynomials are identical or disjoint.  
For developments to Collin's algorithm see for example the Introduction of \cite{BDEMW13}. 

We use \texttt{PL-CAD} for CADs built by projection and lifting and \texttt{RC-CAD} for CADs built via CCDs.  We will discuss a freely available \textsc{Maple} package \textsc{ProjectionCAD} which builds \texttt{PL-CAD}s by utilising routines developed for \texttt{RC-CAD}.  
We continue in Section \ref{SECTION:Why} by describing the motivation for coupling these approaches before explaining the workings of the package in Section \ref{SECTION:How} and describing the current functionality in Section \ref{SECTION:Func}.  Earlier versions of the package can be downloaded alongside \cite{England13a} \cite{England13b}, with the latest version available from the authors.  There are plans for its integration into the \textsc{RegularChains} Library \cite{RC} itself.

\section{Motivation}
\label{SECTION:Why}

\textsc{ProjectionCAD} uses routines in the \textsc{RegularChains} Library to build cells in the lifting phase.  The advantages of utilising the routines are multiple:
\begin{itemizeshort}
\item It avoids many costly algebraic number calculations by using efficient algorithms for triangular decomposition.  
When algebraic numbers are required (as sample points for lower dimension cells) they are represented as the unique root of a regular chain in a bounding box.
\item It ensures \texttt{ProjectionCAD} will always use the best available sub-algorithms in \textsc{Maple}, such as the recently improved routines for real root isolation.
\item It allows \texttt{ProjectionCAD} to match output formats with the \texttt{RC-CAD} algorithms.  In particular, it allows for the use of the sophisticated \emph{piecewise} interface \cite{CDMMXXX09} which highlights the tree-like structure of a CAD.
\end{itemizeshort}
The \textsc{ProjectionCAD} package was developed to implement new theory for \texttt{PL-CAD}, most notably the work in \cite{BDEMW13}, \cite{BDEMW14}, \cite{BDEW13} and \cite{WBDE14}.  More details of the functionality are given in Section \ref{SECTION:Func}.  However, it has also allowed for easy comparison of \texttt{PL-CAD} and \texttt{RC-CAD}, leading to new developments for \texttt{RC-CAD} \cite{BCDEMW14} \cite{EBCDMW14}.  A future aim is identification of problem classes suitable for one approach or the other.

\section{CAD construction in \textsc{ProjectionCAD}}
\label{SECTION:How}

The pseudo code in Algorithm \ref{alg:PLCAD} describes the framework which all algorithms to build CADs within \textsc{ProjectionCAD} follow.  They apply to either polynomials or formulae.  If the former then the CAD produced is sign-invariant for each polynomial.  If the latter then the CAD is such that each formula has constant Boolean truth value on each cell, said to be \textit{truth-invariant} for the formula (\textit{truth table invariant} for the sequence of formulae).  Depending on the algorithm used the user may also have to supply additional information (such as which projection operator to use or which equational constraint to designate \cite{McCallum1999}).  
All algorithms require a specified variable ordering, which can have a significant affect on the tractability of using CAD \cite{BD07}.  We use ordered variables $x_1 \prec \dots \prec x_n$ and say the \textit{main variable} is the highest ordered variable present.

\vspace*{10pt}

\begin{algorithm}[H]
\caption{\texttt{PL-CAD}} 
\label{alg:PLCAD}
\DontPrintSemicolon
\SetKwInOut{Input}{Input}\SetKwInOut{Output}{Output}
\Input{A variable ordering $x_1 \prec \dots \prec x_n$ and $F$ a sequence of polynomials (or quantifier-free formulae).
}
\Output{A CAD of $\mathbb{R}^n$ sign-invariant for the polynomials (or truth invariant for the formulae) $F$; or FAIL if $F$ is not well-oriented. }
\BlankLine
Run the projection phase using an appropriate projection operator on $F$. 
\label{step:Proj} \;
\For{$i=1,\ldots,n$ \label{step:PP1}}{
Assign to $P_i$ the set of projection polynomials with main variable $x_i$. 
\label{step:PP2} \;
}
Set $C_1$ to be a CAD of $\mathbb{R}^1$ formed by the decomposition of the real line according to the real roots of the polynomials in $P_1$.
\label{step:R1} \;
\For{$i=2,\ldots,n$}{
	\For{\emph{each cell} $c \in C_{i-1}$}{
		Check any necessary well-orientedness conditions.
		\label{step:WOCheck} \;
		\eIf{the input is not well oriented}{
			\eIf{dim$(c)=0$}{
				Assign to $L$ a set containing the polynomials in $P_i$ and any (non-constant) minimal delineating polynomials. 
				\label{step:MDP} \;
				}{
				\Return FAIL.
				\label{step:FAIL} \;
				}
			}{
			Set $L := P_i$.
			}
			Set $S_c := \texttt{GenerateStack}(c, L)$.
			\tcp{Apply Algorithm \ref{alg:CADGenerateStack}.} 
			\label{step:GS} \;  
	}
	Set $C_i := \bigcup_c S_c$. 
	\label{step:CollectStacks} \;
}
\Return $C_n$.
\label{step:ReturnRn} \;
\end{algorithm}

\vspace*{10pt}

All algorithms in \textsc{ProjectionCAD} start with a projection phase (step \ref{step:Proj}) which uses a projection operator appropriate to the input to derive a set of projection polynomials.  In steps \ref{step:PP1}-\ref{step:PP2} we sort these into sets $P_i$ according to their main variable.  The remainder of the algorithm defines the lifting phase.  We start by decomposing $\mathbb{R}^1$ into cells according to the real roots of $P_i$ (step \ref{step:R1}) and then repeatedly lift by generating stacks over cells until we have a CAD of $\mathbb{R}^n$.  
All cells are equipped with a \emph{sample point} and a \emph{cell index}.  The index is an $n$-tuple of positive integers that corresponds to the location of the cell relative to the rest of the CAD. Cells are numbered in each stack during the lifting stage (from most negative to most positive), with sectors having odd numbers and sections having even numbers.  Therefore the dimension of a given cell can be easily determined from its index: simply the number of odd indices in the $n$-tuple. 

Before lifting over a cell we first check for the satisfaction of any conditions necessary to conclude the correctness of the theoretical algorithm being implemented (step \ref{step:WOCheck}). 
These conditions are collectively refereed to as the input being \textit{well-oriented} and involve ensuring that projection polynomials are not \textit{nullified} (meaning a polynomial with main variable $x_i$ is not identically zero over a cell in $\mathbb{R}^{i-1}$).  Which polynomials must be checked varies with the algorithm (see \cite{McCallum1998}, \cite{McCallum1999}, \cite{BDEMW13}, \cite{BDEMW14} for details).  If the conditions are not satisfied then an error message is returned in step \ref{step:FAIL}, unless the cell in question is zero-dimensional when correctness can be restored by generating the stack with respect to minimal delineating polynomials (see \cite{Brown2005a}) as well as the projection polynomials in $P_i$ (step \ref{step:MDP}).  Note that input not well-oriented for one operator may be for another, and that Collins' operator is always successful (given sufficient resources).

Building the stack is passed to Algorithm \ref{alg:CADGenerateStack} by step \ref{step:GS}.  The stacks are collected together in step \ref{step:CollectStacks} to form a CAD of $\mathbb{R}^i$ and the final CAD of $\mathbb{R}^n$ returned in step \ref{step:ReturnRn}.  The correctness of Algorithm \ref{alg:PLCAD} follows from the correctness of Algorithm \ref{alg:CADGenerateStack} and the correctness of the various \texttt{PL-CAD} theories implemented proved in their respective papers (for which see the citations in Section \ref{SECTION:Func}).

\vspace*{-10pt}

\subsubsection{Algorithm \ref{alg:CADGenerateStack}}

Stacks are generated following Algorithm \ref{alg:CADGenerateStack}.  It finishes in step \ref{step:RCGS} with a call to  \texttt{RegularChains:-GenerateStack}, an algorithm described in Section 5.2 of \cite{CMXY09} (and implemented in \textsc{Maple}'s \texttt{RegularChains} library).  Algorithm \ref{alg:CADGenerateStack} requires the input be projection polynomials: implying they satisfy the delineability conditions necessary for the cells produced when lifting to have the required invariance condition.  The regular chains algorithm has stricter criteria, requiring in addition that the polynomials \textit{separate above the cell}, meaning they are coprime and squarefree throughout.  Hence Algorithm \ref{alg:CADGenerateStack} must first pre-process to meet this condition.

In steps \ref{step:extract1} and \ref{step:extract2} we simply extract information from the cell $c$ to be lifted over.  We identify those dimensions of the cell which are restricted to a point by consulting the cell index (those indices with even integers) and collect together the equations defining these restrictions in steps $\ref{step:collect1}-\ref{step:collect2}$.  Note that there is no ambiguity in the ordering of the polynomials in $E$ since a regular chain is defined by polynomials of different main variables \cite{ALM99}.
If the cell is of full dimension then there is no need to process since the polynomials are delineable and taken from a squarefree basis.  Otherwise, we process using Algorithms \ref{alg:MakeCoprime} and \ref{alg:MakeSquareFree} in steps \ref{step:MkCP} and \ref{step:MkSF}.  
The restriction is identified using a regular chain $\hat{rc}$ (step \ref{step:rcdash}) together with the original bounding box.  We can be certain that $\hat{E}$ defines a single regular chain since the equations were extracted from one.

\vspace*{-10pt}

\subsubsection*{Algorithm \ref{alg:MakeCoprime}}

In order to make the polynomials coprime we use repeated calls to a triangular decomposition algorithm in step \ref{step:Tri} (described in \cite{MorenoMaza2000} and part of the \texttt{RegularChains} Library).  Given lists of polynomials $L_1$ and $L_2$ and a regular chain, it returns a decomposition of the zeros of $L_1$ which are also also zeros of the regular chain but not zeros of $L_2$.  We use $\hat{rc}$ for the regular chain, so we work on the restriction, and build up a list of coprime polynomials by ensuring existing ones ($L_2$) are not zeros in decompositions of the next one ($L_1$).  
Each time the decomposition is a list of either regular chains or \textit{regular systems} (a regular chain and an inequality regular with respect to the chain \cite{Wang2000}).  

We consider each of these components in turn.  If the main variable is lower then the solution is discarded.  
Otherwise we check if the component has a solution compatible with the sample point for the cell (as it may be a solution of $\hat{rc}$ other than one isolated by $bb$).  This means isolating the real solutions (of the component excluding the top dimension) and refining their bounding boxes until they are either within $bb$ or do not intersect at all.  It is achieved using the \texttt{RealRootIsolate} command in the \texttt{RegularChains} Library (see \cite{BCLM09}).
Finally if the component passes these tests then the polynomial in the main variable is extracted and added to the set returned from Algorithm \ref{alg:MakeCoprime} in steps $\ref{step:CPadd1}-\ref{step:CPadd2}$.  

\vspace*{-10pt}

\subsubsection*{Algorithm \ref{alg:MakeSquareFree}}

In order to make the polynomials squarefree we use repeated calls to an algorithm which does this modulo a regular chain ($\hat{rc}$: so that we are working on the restriction).  It is an analogue of Musser's \cite{Musser1975} with the gcd calculations performed modulo the regular chain as described in \cite{LMP09}.
It assumes the polynomial is regular modulo the chain and so we first test for this.  If not regular (the leading coefficient vanishes) then we consider the \texttt{tail} (polynomial minus the leading term) in step \ref{step:tail}, if still in the main variable.
The output of the factorization is either: $rc$ and a list of polynomials forming a squarefree decomposition of $p$ modulo $rc$; or a list of pairs of regular chains and squarefree decompositions where the regular chains are a decomposition of $rc$.  In the latter case only one will be relevant for the root isolated by $bb$ and we identify which using the \texttt{RealRootIsolate} command, similarly to Algorithm \ref{alg:MakeCoprime}.

\vspace*{5pt}

\begin{algorithm}[H] 
\caption{\texttt{GenerateStack}}
\label{alg:CADGenerateStack}
\DontPrintSemicolon
\SetKwInOut{Input}{Input}\SetKwInOut{Output}{Output}
\Input{A cell $c$ from a CAD of $\mathbb{R}^{k}$ and a set $P$ of projection polynomials in $x_1 \dots x_{k+1}$ (part of a squarefree basis).
}
\Output{A set of cells $\mathcal{S}$ of $\mathbb{R}^{k+1}$ comprising a stack over $c$.  The polynomials in $P$ are sign-invariant on each cell of $\mathcal{S}$.
}
\BlankLine
Set $I$ and $sp$ to be the cell index and sample point of $c$.
\label{step:extract1}\;
Set $rc$ and $bb$ to be the regular chain and bounding box encoding $sp$. 
\label{step:extract2} \;
Set $E$ to be the set of $k$ polynomials whose zeros define $rc$, ordered by increasing main variable.
\label{step:collect1} \;
Set $\hat{E} := \{ \, \}$.\;
\For{$i=1,\ldots,k$}{
	\If{the $i$'th integer in $I$ is even}{
		Add the $i$th polynomial in $E$ to $\hat{E}$.
		\label{step:collect2} \;  
	}
}
\If{$\hat{E} \neq \{ \, \}$}{ 
	Set $\hat{rc}$ to be the regular chain formed by $\hat{E}$.
	\label{step:rcdash} \;
	$P := \texttt{MakeCoprime}(P, \hat{rc}, c)$.
	\label{step:MkCP}
    \tcp{Apply Algorithm \ref{alg:MakeCoprime}.}
	$P := \texttt{MakeSquareFree}(P, \hat{rc}, c)$.
	\label{step:MkSF}
    \tcp{Apply Algorithm \ref{alg:MakeSquareFree}.}
}
$\mathcal{S} := \texttt{RegularChains:-GenerateStack}(c,\hat{P}, k+1)$ 
\label{step:RCGS} 
\tcp{From \cite{CMXY09}.}
\Return $\mathcal{S}$. \;
\end{algorithm} 

\newpage

\begin{algorithm}[H]
\caption{\texttt{MakeCoprime}}
\label{alg:MakeCoprime}
\DontPrintSemicolon
\SetKwInOut{Input}{Input}\SetKwInOut{Output}{Output}
\Input{
	A set of polynomials $P$, 
	a regular chain $\hat{rc}$ and a cell $c$.
}
\Output{
	A set of polynomials $\hat{P}$ which describe the same set of varieties, but which are coprime over $c$.\\
}
\BlankLine
Set $\hat{P} = \{ \, \}$.\;
\For{polynomial $p \in P$}{
	$T := \texttt{Triangularize}([p],\hat{P},\hat{rc})$.
	\label{step:Tri}
	\tcp{From \cite{MorenoMaza2000}.}
	\For{component $\mathcal{C}$ of $T$}{
		\If{${\rm mvar}(\mathcal{C}) \neq {\rm mvar}(p)$}{
			\texttt{next} $\mathcal{C}$ 
		}
		\If{$\mathcal{C}$ has a zero compatible with the sample point of $c$}{
			Set $\hat{p}$ to be the polynomial in $\mathcal{C}$ with main variable $p$.
			\label{step:CPadd1} \;
			Add $\hat{p}$ to $\hat{P}$. 
			\label{step:CPadd2} \;	
		}	
	}
}
\Return $\hat{P}$. \;
\end{algorithm}

\vspace*{10pt}

\begin{algorithm}[H]
\caption{\texttt{MakeSquareFree}} 
\label{alg:MakeSquareFree}
\DontPrintSemicolon
\SetKwInOut{Input}{Input}\SetKwInOut{Output}{Output}
\Input{
	A set of polynomials $P$, 
	a regular chain $\hat{rc}$ and a cell $c$.
}
\Output{
	A set of polynomials $\hat{P}$ which describe the same set of varieties, but which are each squarefree over $c$.
}
\BlankLine
Set $\hat{P} = \{ \, \}$.\;
\While{P is not empty \label{step:startwhile}}{
    Remove a polynomial $p$ from $P$. \;
		\If{$p$ is not regular over $\hat{rc}$}{
			Set $\hat{p} = \texttt{tail}(p)$
			\label{step:tail} \;
			\If{${\rm mvar}(\hat{p}) = {\rm mvar}(p)$}{
				Add $\hat{p}$ to $P$ and \texttt{continue} from step \ref{step:startwhile}.
				\label{step:SFadd1} \; 
			}
		}
    $T := \texttt{SquarefreeFactorization}(p,\hat{rc})$
    \label{step:SFF} \;
	Select $\mathcal{C}$ as the component in $T$ compatible with the sample point of $c$. \; 
	Set $\hat{p}$ to be the product of polynomials in the decomposition in $\mathcal{C}$. \;
	Add $\hat{p}$ to $\hat{P}$.
	\label{step:SFadd2} \; 	
}
\Return $\hat{P}$.\;
\end{algorithm}

\section{Functionality of ProjectionCAD}
\label{SECTION:Func}

We finish by listing  some of the functionality of within \textsc{ProjectionCAD}, focusing on aspects not usually found in other CAD implementations:
\begin{itemize}
\item Sign-invariant CADs can be built using the Collins \cite{Collins1975} or McCallum \cite{McCallum1998} projection operators.
\item CADs can be built with the stronger property of \textit{order-invariance} (where each polynomial has constant order of vanishing on each cell) \cite{McCallum1998}.
\item \textit{Equational constraints} (ECs) are equations logically implied by the formula. They can be utilised via McCallum's reduced projection \cite{McCallum1999} and a more efficient lifting phased (detailed in Section 5 of \cite{BDEMW14}). 
\item TTICADs can be built for sequences of formulae, making use of ECs in each \cite{BDEMW13} \cite{BDEMW14}.  TTICAD can be both a desired structure for applications \cite{EBDW13} and an efficient way to build a truth-invariant CAD (allowing savings from ECs for conjunctive sub-formulae, not ECs of the whole formula).
\item Minimal delineating polynomials \cite{Brown2005a} are built automatically, avoiding unnecessary failure declarations (which can occur in \textsc{Qepcad}).  See \cite{England13a} for an example of this.
\item User commands for stack generation and the construction of \textit{induced CADs} (a CAD of $\mathbb{R}^i, i<n$ produced en route to a CAD of $\mathbb{R}^n$), allowing for easy experimentation with the theory.
\item \textit{Layered CADs} contain cells of only a certain dimension or higher. They can be produced (more efficiently than a full CAD) \cite{WBDE14}. 
\item \textit{Variety CADs} contain only those cells that lie on the variety defined by an EC.  They can be produced (more efficiently than a full CAD) \cite{WBDE14}. 
\item Layered and manifold TTICADs as well as layered-manifold CADs can be produced \cite{WBDE14} (combining the savings from the different theories). 
\item Heuristics are available to help with choices such as variable ordering, EC designation and breaking up parent formulae for TTICAD \cite{BDEW13}.
\end{itemize}
Details can be found in the citations above and the technical reports \cite{England13a}, \cite{England13b}.

\subsection*{Acknowledgements}
This work was supported by the EPSRC (grant number EP/J003247/1).  We thank the developers of the \textsc{RegularChains} Library, especially Changbo Chen and Marc Moreno Maza, for access to their code and assistance working with it.


\end{document}